\documentclass[pdflatex,sn-mathphys-num]{sn-jnl}
\usepackage{graphicx}%
\usepackage{multirow}%
\usepackage{amsmath,amssymb,amsfonts}%
\usepackage{amsthm}%
\usepackage{mathrsfs}%
\usepackage[title]{appendix}%
\usepackage{xcolor}%
\usepackage{textcomp}%
\usepackage{manyfoot}%
\usepackage{booktabs}%
\usepackage{algorithm}%
\usepackage{algorithmicx}%
\usepackage{algpseudocode}%
\usepackage{listings}%
\usepackage{caption}

\begin{document}

\title{Continuous and discrete compartmental models for infectious disease}

\author[1]{\fnm{Gustavo A.} \sur{Sousa}}
\equalcont{These authors contributed equally to this work.}
\author[2]{\fnm{Diogo L. M.} \sur{Souza}}
\equalcont{These authors contributed equally to this work.}
\author*[3]{\fnm{Enrique C.} \sur{Gabrick}}\email{ecgabrick@gmail.com}
\equalcont{These authors contributed equally to this work.}
\author[2]{\fnm{Patr\'icio D. C.} \sur{dos Reis}}
\equalcont{These authors contributed equally to this work.}
\author[2]{\fnm{Lucas E.} \sur{Bentivoglio}}
\equalcont{These authors contributed equally to this work.}
\author[2,4]{\fnm{Antonio M.} \sur{Batista}}
\equalcont{These authors contributed equally to this work.}
\author[2,3,4]{\fnm{Jos\'e D.} \sur{Szezech Jr.}}
\equalcont{These authors contributed equally to this work.}

\affil[1]{ Department of Physics, State University of Ponta Grossa, 84030-900, Ponta Grossa, PR, Brazil.}
\affil[2]{Graduate Program in Science, State University of Ponta Grossa,
84030-900, Ponta Grossa, PR, Brazil}
\affil[3]{Institute of Physics, University of São Paulo, 05315-970, São Paulo, SP, Brazil}
\affil[4]{Department of Mathematics and Statistics, State University of Ponta
Grossa, 84030-900, Ponta Grossa, PR, Brazil}

\abstract{
The study of infectious disease propagation is essential for understanding 
and controlling epidemics. One of the most useful tools for gaining 
insights into the spread of infectious diseases is mathematical modelling. In terms of 
mathematical epidemiology, the main models are based on compartments, 
such as Susceptible--Infected (SI), Susceptible--Infected--Recovered (SIR), 
and Susceptible--Exposed--Infected--Recovered (SEIR). These models offer 
mathematical frameworks for representing the proliferation dynamics of 
various diseases, for instance flu and smallpox. In this work, we explore 
these models using two distinct mathematical approaches, Cellular 
Automata (CA) and Ordinary Differential Equations (ODEs). They  
are able to reproduce the spread dynamics of diseases with their own 
individuality. CA models incorporate the local interaction among 
individuals with discrete time and space, while ODEs   
provide a continuous and simplified view of a disease propagation in 
large and homogeneous populations. By comparing these two approaches, we 
find that the shape of the curves of all models is similar for both 
representations. Although, the growth rates differ between CA and ODE. 
One of our results is to show that the CA yields a power-law growth, while the ODE 
growth rate is well-represented by an exponential function. 
{ Furthermore, a substantial contribution of our work is using a hyperbolic tangent to fit the initial growth of infected individuals for all the considered models. Our results display a strong correlation between simulated data and adjusted function. We mainly address this successful result by the fact that the hyperbolic function captures both growing: the power-law (when considered the first terms of infinite sums) and combinations of exponential (when the hyperbolic function is written via exponential).}
{ Therefore}, our work shows that when 
modelling a disease the choice of mathematical 
representation is crucial, in particular to model the onset of an
 epidemic.
}

\keywords{Cellular automata, Epidemiology, Differential Equations.}

\maketitle


\section{Introduction}
Infectious disease spread is a worldwide problem that has affected the humanity 
throughout the history \cite{KEELING}. Understanding the spread of illness 
is an important topic for the modern science, improving the forecast 
and control methods \cite{ANDERSON}. One powerful tool to investigate 
the disease spread is the mathematical models \cite{Bjornstad}. The investigation 
can be done via statistical modelling \cite{Cummings2009}, Ordinary Differential 
Equations (ODEs) \cite{Manchein2020}, Partial Differential Equations \cite{Yamazaki2018}, Cellular 
Automaton (CAs) \cite{White2007}, Machine Learning \cite{Wang2020}, among others 
\cite{Hoertel2020, Satorras2015, Allen1994, Allen2000}. In some cases, 
specifically in ODE and CA, the modelling is mostly made based on 
compartmental models \cite{KEELING}.

Compartmental epidemiological models have their origins in attempts to 
quantify and predict the spread of infectious diseases. One of the most 
influential models was proposed by William Hamer in 1906, using the 
concept of ``mass action'' transmission \cite{Hamer}. In this model, the 
rate of new cases is proportional to the product of the health and 
the sick individuals. However, it was the model proposed by Kermack 
and McKendrick \cite{KERMACK}, in 1927, that laid the foundation for modern 
compartmental models. 

These models compartmentalise the host population according 
to the status of infection. The main compartments are Susceptible ($S$), 
Exposed ($E$), Infected ($I$) and Recovered ($R$) states \cite{BATISTA}. 
$S$ compartment is responsible for holding healthy individuals who can contract 
the infection, $E$ compartment contains the individuals who are infected but not 
infectious, namely latent period \cite{Li1995}. The infectious agents 
are in the $I$ compartment. After the individuals pass through the infectious 
stage, they evolve to $R$. 
Combinations of these compartments and inclusion of new ones lead to several models. 
In this work, we focus on Susceptible--Infected (SI), 
Susceptible--Infected--Recovered (SIR) and
Susceptible--Exposed--Infected--Recovered (SEIR) models. The first model 
is adequate for diseases that confer a lifelong infection, such as 
AIDS \cite{Cai2009}. The second model is adequate for diseases in which the 
healthy individual becomes infected and, after a time interval, acquires a permanent 
immunity, for instance measles \cite{Bjornstad2002}. The SEIR  
is employed to describe illness with latent period, and one recent 
example is COVID-19 \cite{MOHAJAN}.
These models can be described in different mathematical frameworks. In this 
work, we analyze the CA and ODE and compare their results. 

CAs are mathematical models capable of simulating complex biological, 
chemical, and physical processes \cite{NEUMANN}. They were proposed by Stanislaw Ulam and 
John von Neumann during the 1940s. The CA model is defined as a grid of cells that evolve 
according to local rules based on the states of their nearest neighbors \cite{WOLFRAM3}. 
Later, the concept was expanded, most notably with John Conway's ``Game of Life'', which used 
CA to demonstrate complex behaviors emerging from simple rules \cite{GARDNER1}.

In mathematical epidemiology, CAs have been used to model the spread of infectious 
diseases in discrete and spatially distributed populations \cite{SIRAKOULIS}. 
Their relevance grew during the 1980s and 1990s with 
research focusing on the influence of spatial structure and human mobility 
on disease dissemination \cite{YAKOWITZ, FUENTES}. CA has been applied to 
study the spread of HIV \cite{Zorzenon2001}, dengue \cite{Pereira2018}, and influenza outbreaks \cite{Catherine2005}, 
allowing the simulation of hypothetical scenarios and the integration of 
real data for more accurate predictions and strategic control \cite{MUGNAINE}.

In the study of epidemiology, ODEs are employed to model the 
dynamics of infectious disease transmission in homogeneous populations, 
enabling the analysis of factors such as transmission rate, recovery rate, 
and the impact of interventions, providing effective approaches for 
infectious disease control \cite{CHAVEZ, BEIRA}.

In this work, we study the compartmental models SI, SIR, and SEIR  
using CA and ODE. 
By comparing the results, we aim to identify and analyse the similarities 
and differences between these two  approaches. Our results show 
that the CA and ODE exhibit distinct growth rate for the infected 
populations in the models studied. The growth of the infected population 
in the CA model follows a power-law function, while in the ODE model 
is better represented by an exponential function. Our study highlights 
the unique characteristics of each mathematical approach, enabling future 
studies to choose the most appropriate method for accurately representing 
experimental data. { Moreover, a key finding is that the hyperbolic tangent is a strong candidate to fit the initial growth of infected individuals for all the considered models in both representations. This occurs mainly due to the fact that the hyperbolic tangent encompasses power-law growth (in the first terms of infinite sums) and exponential representations.}

This work is organized as following: In Section \ref{methodology}, we 
present the mathematical framework of CA 
and ODE. Thereafter, the SI model is 
discussed in Section \ref{si_section}. Results concerning to SIR and 
SEIR model are present in Sections \ref{sir_section} and \ref{seir_section}, 
respectively. Finally, we draw our conclusions in Section \ref{discussion_section}.

\section{Methodology}\label{methodology}
Cellular automata are considered discrete 
representations of partial differential equations \cite{WOLFRAM3}. 
They are mathematical models characterized by discrete space, time, and 
state variables \cite{ILACHINSKI}. A model based on CA framework can 
have $n$ space dimensions. In this work, we consider $n=2$, where $n$ 
is the space dimension,  and a square 
grid compose of $M \times M$ identical cells ($C_{i,j}$). The time evolution 
depends on pre-defined rules and the neighborhood. The rules are defined 
according to the problem to be modelled and the neighborhood is 
defined by taking the closest neighbors of $C_{i,j}$. In this work, we 
consider a two space dimension where the neighbors belongs to the von Neumann 
neighborhood, i.e., the set of neighbors in 
the four closest cells \cite{MUGNAINE}. The time step is defined when the 
transition rules are applied to the whole 
grid, simultaneously. 

In this work, for each epidemiological model, we assume that each cell 
in the CA represents one individual. Our grid consist of $M \times M$ cells, 
where $M=100$, implying in a population equal to $N = 10000$, 
and we employ the von Neumann neighbourhood with periodic boundary 
conditions. The transition rules specific to each compartmental model 
are detailed in their respective sections. 
The position of the infected cells in the grid are randomly 
selected. Therefore, the CA model considered is probabilistic,  the epidemic 
curves are an average of 100 independent simulations.

Since their development, ODEs have been applied across various fields of 
knowledge. In physics, they are used to describe the behaviour of dynamic 
systems, such as neuronal modelling \cite{SOUZA2024},  
pendulum oscillations \cite{ROSENBLUM}, fluid dynamics \cite{KIKUCHI}, and others. 
In epidemiological studies, ODEs have been used to model the dynamics of 
infectious disease transmission in homogeneous populations \cite{HETHCOTE}. 
They allow the analysis of factors such as transmission rates, recovery 
rates, and the impact of interventions, providing an effective approach 
for developing disease control strategies \cite{DIEKMANN,TIAN}.

In this work, to numerically integrate the differential equations, 
we use the 4th-order Runge-Kutta method with a fixed step equal to $h=0.01$.

\section{SI model}\label{si_section}
\begin{figure}[htb]
\centering
\includegraphics[scale = 0.2]{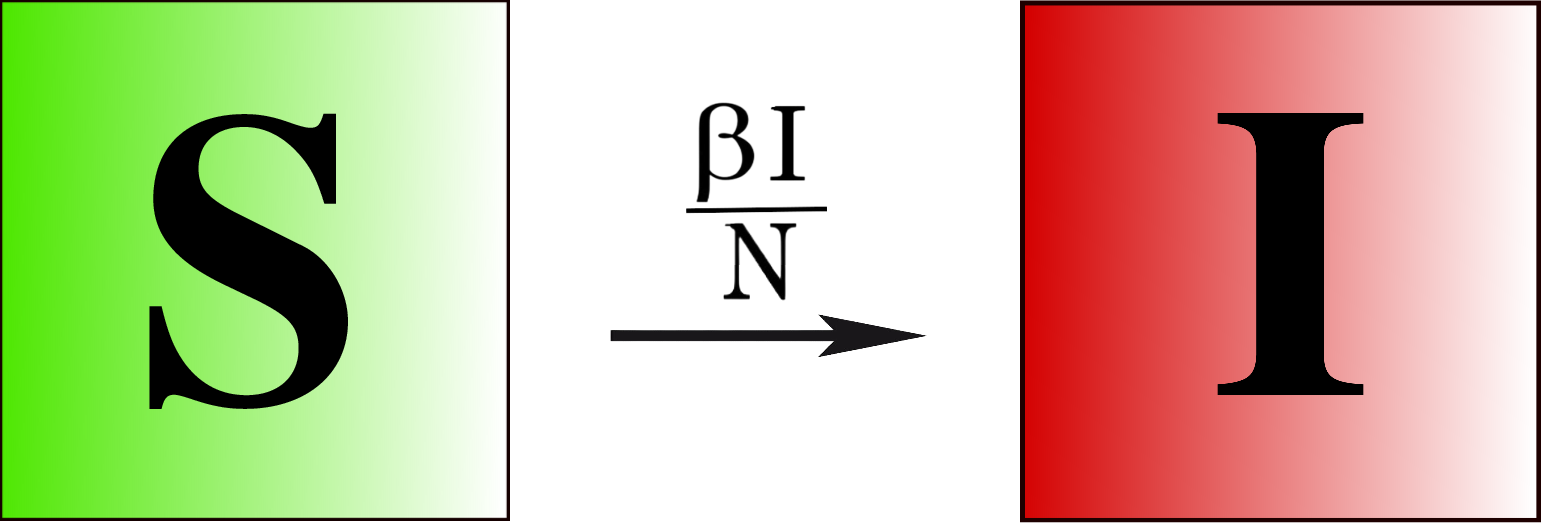}
\caption{Compartmental diagram of the SI model.}
\label{fig1}
\end{figure}

One of the simplest compartmental models extensively used is the SI. 
This model describes the spread of diseases where the $I$ individuals have lifelong  
infection, such as AIDS \cite{Dalal2007} or mosquitoes infected  by dengue \cite{Aguirre2023}. 
Figure \ref{fig1} shows the interaction between the $S$ and 
$I$ compartments, where the arrow indicates the flow of individuals. 
Above the arrow, we include the notation $\beta I/N$, which represents the 
force of infection, that define the infection contraction of each $S$ individual \cite{KEELING}, 
and $\beta$ is the product between transmission probability and contact rates.
In this way, the rate in which new infected individuals are produced is 
$\beta S I /N$.

Firstly, we simulate the SI model using CA approach. During each iteration, 
if one or more neighbors are $I$, 
each $S$ can evolve to $I$ with a rate $\beta$. Once a cell becomes 
$I$, it remains in that compartment for the rest of the simulation. 
This approach allows an analysis of the spatial and temporal dynamics 
of the infection, which is useful for understanding the spread patterns. 
We consider states set $U = \{S,I\} = \{1,2\}$. 
Then, the update rules of the each cell $C_{i,j}(t)$ (Susceptible or Infected) are summarized as follow:
\begin{enumerate}
\item The susceptible cells verify the presence of 
neighbour infected cells, than each adjacent infected cell  
infects the susceptible cell with a probability $\beta$.
\item Once the susceptible cell is infected, it evolves 
to the infected state  and stays there trying to contaminate
S adjacent cells.
\end{enumerate}

\begin{figure}[htb]
\centering
\vspace{5pt}
\includegraphics[scale = 0.4]{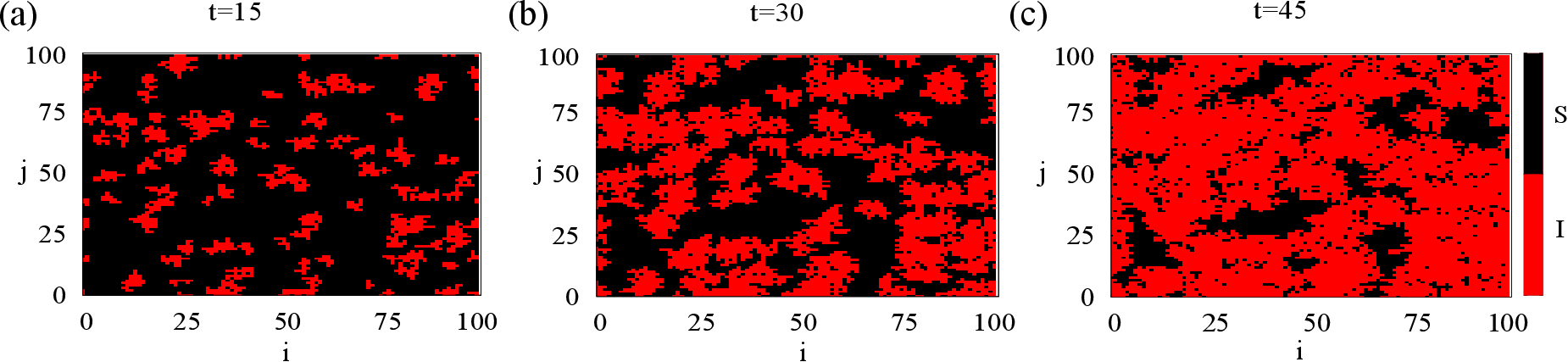}
\caption{Time and spatial evolution of the SI model via CA. 
The black color displays the susceptible and red the infected individuals.}
\label{fig2}
\end{figure}

By implementing the rules with parameters listed in Table \ref{tab_2}, 
we show the spatial evolution of SI model in Fig. \ref{fig2}, in CA 
framework. 
{ It is worth mentioning that the parameter values are arbitrary, since we are not modeling any real disease, but rather exploring fundamental relationships of both formulations (ODE and CA). However, once selected the parameters values, they need to be equal for both approaches, including the fraction of initial individuals.  }
Figures \ref{fig2}(a), \ref{fig2}(b), and \ref{fig2}(c) 
are snapshots at $t=15$, 30, and 45, respectively. In these results, 
and in the rest of the paper, we represent $S$ cells by black colour and $I$ 
by red. For $t=15$ (Fig. \ref{fig2}(a)), we observe the emergence of many  
infection focus, which are centred in the infected initial condition. 
As $t$ advances, each aggregation grows in the grid. In Fig. \ref{fig2}(c),  
we select the snapshot when $t=45$, for in this a significant 
part of the grid is red. 

Nonetheless, the equilibrium is reached when 
all the cells become $I$; for our parametric configuration, this occurs 
at $t \approx 100$. It is important to mention that these 
snapshots are for one random initial condition, if the initial infected cells 
are disposed in other position, the snapshots assume different shape. 
A major advantage of CA is the ability to capture 
spatial variations, thereby representing the rapid infection of densely 
populated areas.

\begin{table}[hbt]
    \caption{Parameters for the SI model considering CA and 
    ODEs. The initial 
    condition for $S$ is $S(0) = N - I(0)$.}
    \centering
    \begin{tabular}{|c|c|c|c|}
    \hline
    Parameter & Values (CA) & Values (ODE) \\
    \hline
    $N$     & 10000             & 1  \\
    $\beta$ & 0.1               & 0.1\\
    $I(0)$  & 100 (0.01$N$)     & 0.01\\
    \hline
    \end{tabular}\label{tab_2}
    \end{table}

Next, we implement the SI through ODEs. The equations  
represent the change rate of $S(t)$ and $I(t)$ compartments:
\begin{eqnarray}
    \frac{dS}{dt} &=& -\beta \frac{SI}{N}, \label{si_eq1}\\
    \frac{dI}{dt} &=& \beta \frac{SI}{N}, \label{si_eq2}
\end{eqnarray}
where $N = S + I$ is the total population \cite{BATISTA}. Using 
this constraint the model can be uncoupled and solved analytically. Considering 
$S = N - I$ in Eq. (\ref{si_eq2}), we get
\begin{equation}
\frac{dI}{dt} = \frac{\beta}{N}I(N-I),
\end{equation}
where the solution for $I$ is given by 
\begin{equation}
I(t) = \frac{I_0 N e^{\beta t}}{N + I_0(e^{\beta t} - 1)}, \label{sol_si}
\end{equation}
and for $S(t) = N - I(t)$. The $I$ number grows as an exponential function 
which depends on $\beta$. If $S(0) \approx N$ and $I(t) \ll N$, a rapid spread of the disease 
occurs according to an exponential function dependent on $\beta$. Although 
we keep the capital notation in the equations for ODEs approaches, 
the epidemic curves are normalised, i.e., 
$s = S/N$, $i = I/N$, and so on.  

As the disease spreads, the number of susceptible individuals decreases, 
going to zero, while the number of infected individuals reaches a 
plateau close to the total number of individuals in the population. 
Eventually, the number of new cases stabilizes, and all susceptible 
individuals become infected. 

\begin{figure}[htb!]
\centering
\includegraphics[scale = 0.2]{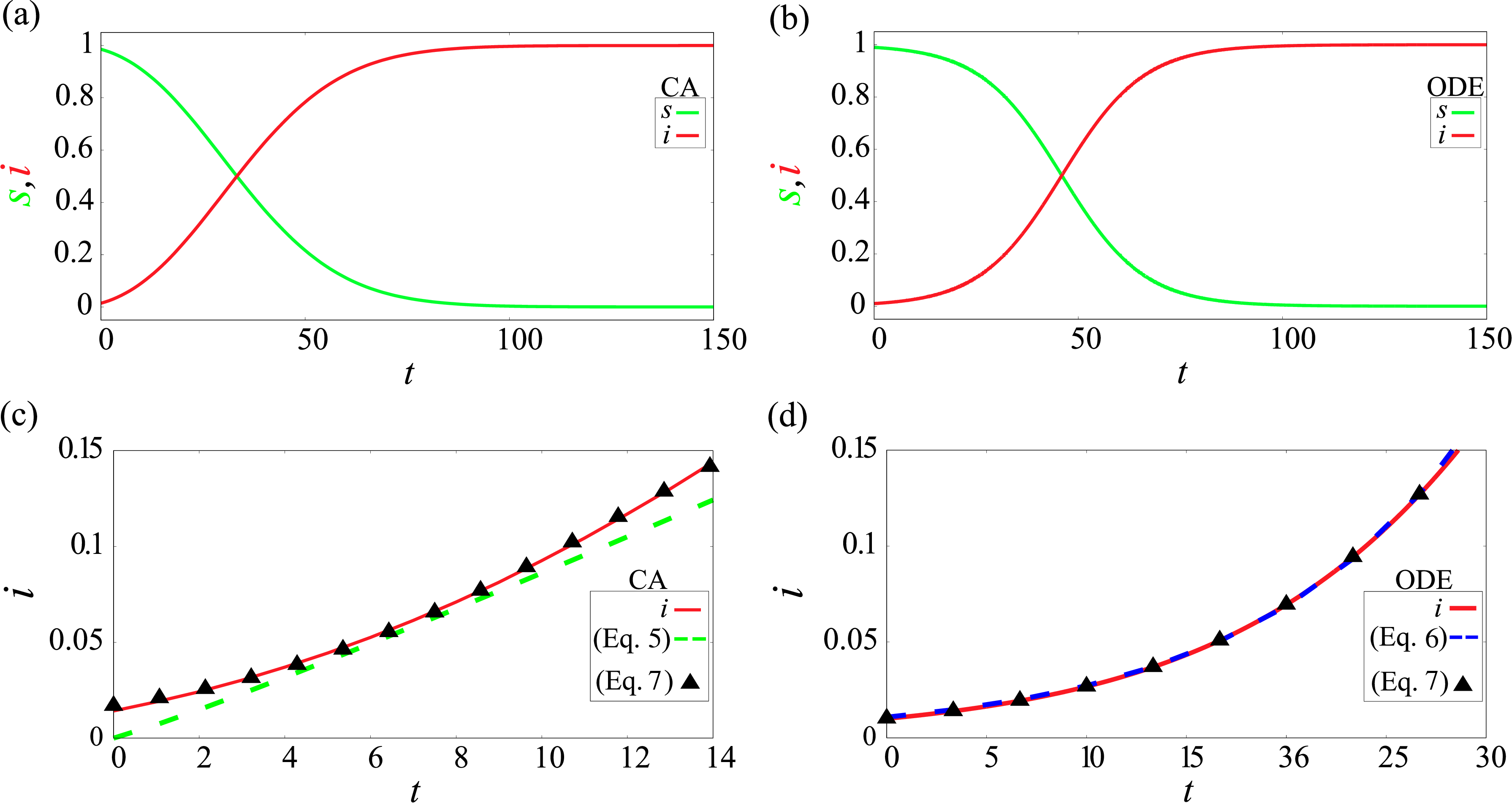}
\caption{Time evolution of $s(t)$ (green line) and $i(t)$ (red line) in  the panels (a) and (b), for 
CA and ODE, respectively.
The panels (c) and (d) display the respective infected curve for both approaches, where  
the red line is for $i$, the green dotted line is the adjustment 
using Eq. (\ref{fit_CA}), the black triangle is 
the fitting using Eq. (\ref{fit_tanh}), and the blue dotted line is the 
adjustment given by Eq. (\ref{fit_ODE}).}
\label{fig3}
\end{figure}

With both mathematical formulation (CA and ODE) for SI 
model, we implement numerical solutions for both cases, the parameters for CA and ODE are described 
in Table \ref{tab_2}. From the CA model, the population in 
$S$ compartment starts to move to $I$ until this compartment to be filled  (Fig. \ref{fig3}(a)). 
The same dynamics is observed in ODE framework. Taking the limit 
$t \rightarrow \infty$ in Eq. (\ref{sol_si}) we get $(S,I)=(0,N)$, 
showing that all the population moves from $S$ to $I$. This is not necessarily  
true for the next analyzed models. A numerical simulation corroborates this 
analysis, as shown in Fig. \ref{fig3}(b), where we observe the changes 
between the compartments. 

The solutions displayed in Fig. \ref{fig3}(a) and Fig. \ref{fig3}(b) have 
a similar behaviour. However, when we analyse the increase of infected 
individuals for short times, we can infer the differences, as exhibited in  
Figs. \ref{fig3}(c) and \ref{fig3}(d). In CA,  
the decrease in susceptible and the increase in infected individuals is slower 
compared to the ODE formulation (Fig. \ref{fig3}(b)). To compare the 
initial growth in both formulations, we propose a 
power-law for CA, 
\begin{equation}
 I_{\rm CA}(t) = Kt^{B}, \label{fit_CA}
\end{equation}
for ODE framework: 
\begin{equation}
 I_{\rm ODE}(t) = K \exp{(B t)}, \label{fit_ODE}
\end{equation}
and to compare both growing, we consider a hyperbolic tangent:
\begin{equation}
\widehat{I} (t) = \alpha \left(1 + {\rm tanh}(\xi t + \phi)\right), \label{fit_tanh}
\end{equation}
where $\alpha \in [0,1]$ is the amplitude, $\xi$ is the growing rate, 
and $\phi$ is a phase. { $\widehat{I} (t)$ corresponds to the individuals, $\rm tanh$ and $\phi$ are dimensionless, and $\xi$ has $1/{\rm time}$ unity. In this way, $\alpha$ is the number of individuals who become sick in a time interval, i.e., an incidence factor. }

The initial time evolution of $i$ curve (until $I = 0.15N$) for both 
approaches is displayed in Fig. \ref{fig3}(c) and \ref{fig3}(d). In the panel  
(c), the red line is for $i$, while the green dotted line is the adjustment 
using Eq. (\ref{fit_CA}) for $K = 0.007 \pm 0.001 $ and 
$B = 1.09 \pm 0.05$, with $R^2 = 0.9966$; and the black triangle is 
the fitting using Eq. (\ref{fit_tanh}), for 
$\alpha = 0.115 \pm 0.005$, $\xi = 0.108 \pm 0.003$, $\phi = -1.27 \pm 0.01$, 
and $R^2 = 0.9998$. Both curves describe the initial grow 
of $i$ individuals. 

We conduct a similar analysis for the ODE formulation (Fig. \ref{fig3}(d)). 
For Eq. (\ref{fit_ODE}), we obtain   
$K = 0.01076 \pm 0.00001$ and $B = 0.09262 \pm 0.00004$, 
and $R^2 = 0.99987$, shown by the blue dotted line; and for Eq. (\ref{fit_tanh}),  
$\alpha = 0.500 \pm 0.003$, $\xi = 0.05000 \pm 0.00003$, $\phi = -2.298 \pm 0.003$, 
and $R^2 = 0.99998$, exhibited by the black triangle. 

In the analysed cases, we obtain a significant 
$R^2$, for all the fits, { but} we note that the infection modelled by CA 
grows as a power-law, while via ODEs follows an exponential. 
A hyperbolic tangent is able to fit both grows. 
For SI this approach works for the whole time of $i(t)$, { however}, 
for other models which present a bell shape of infection, this function 
is adequate just for short and medium times, i.e., times before the peak. 
For these parametric 
configurations and analysing the adjustments, we observe that the infected curve grows faster for CA than 
ODE. This difference in infection speed also arises due to the ODE 
model is treating the population as a continuous medium, ignoring spatial 
interactions between individuals.

\section{SIR model}\label{sir_section}
\begin{figure}[htb!]
\centering
\includegraphics[scale = 0.14]{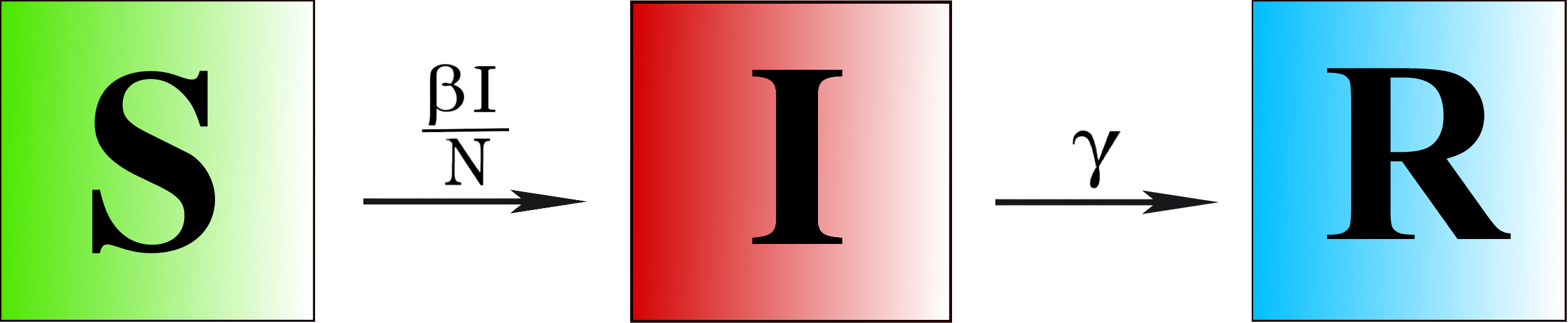}
\caption{Diagram of the compartments for the SIR model. $\beta$ describes the effective rate 
         of infection and $\gamma$ represents the recovery rate.}
\label{fig4}
\end{figure}

The SIR model describes the transition from an 
infected state to a recovered one. It is able to mimic spread of diseases 
where individuals obtain permanent immunity 
after the infectious period, for instance measles \cite{Bjornstad} 
and smallpox \cite{Kuddus}. Figure \ref{fig4} 
shows the interactions between the compartments in SIR. 
We consider a new parameter $\gamma$, which is the recovery rate and 
$1/\gamma$ gives the average infectious time. 

We set up a CA formulation for the SIR model, where the transition 
between different compartments is governed by infection probability $\beta$ 
and recovery probability $\gamma$. The set of states 
is $U = \{S,I,R\} = \{1,2,3\}$. The $S$ cell can become $I$ with probability 
$\beta$ if it has at least one infected neighbor. After a cell becomes 
infected, it recovers with a probability $\gamma$. We do not consider 
mortality due to the disease, { then} after a long simulation time, the 
entire population will be in $R$ compartment. The transition rules of 
SIR are similar to those of SI, the only difference is the addition 
of the transition to $R$ state. A summary of the state transition rules for the SIR model is
\begin{enumerate}
	\item An infected cell can contaminate an adjacent 
	$S$ cell with probability $\beta$. If there are multiple 
	infected cells around a susceptible cell, each $I$ cell contaminates the $S$ cell   
	with probability $\beta$.
	\item When a $S$ cell is infected by one of its neighbours, 
	it evolves to the infected state.
	\item An infected cell can recover from the pathology 
	with probability $\gamma$. Upon recovery, the infected cell goes to the $R$ state and remains there.
\end{enumerate}

\begin{figure}[htb!]
\centering
\includegraphics[scale = 0.4]{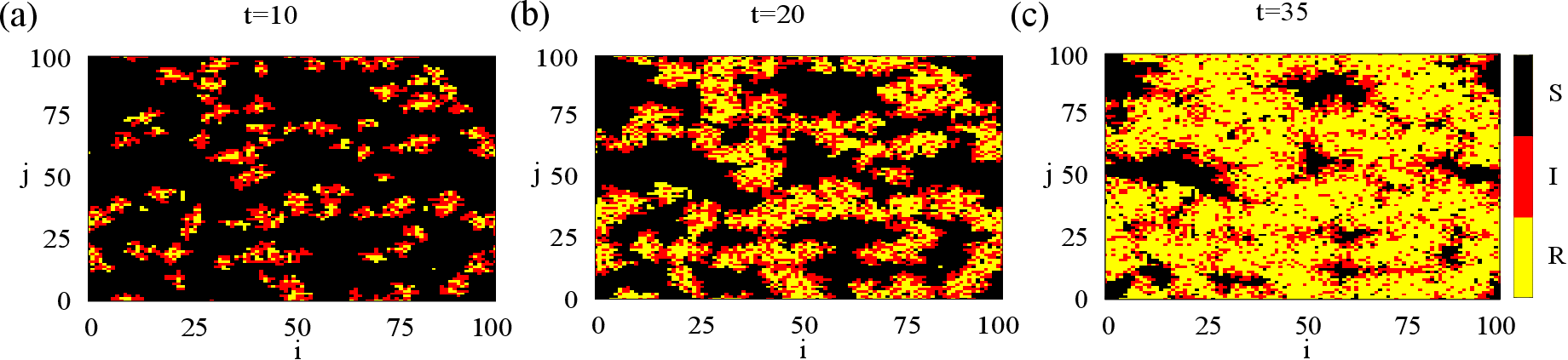}
\caption{Time evolution of CA for SIR for (a) $t=10$, (b) $t=20$, and (c) $t=35$. Black colour represents the susceptible, 
         red colour the infected and yellow the recovered individuals.}
\label{fig5}
\end{figure}

Figure \ref{fig5} shows the spatial and temporal evolution of the SIR based 
on CA formulation with parameters described in Table \ref{tab_4}. At the 
beginning of the simulation (Fig. \ref{fig5}(a)), 
there are only a few infected and recovered cells. Over time, 
there is a wave of $I$ transmitting the disease to $S$, where the growing is like 
a circle centred in the initial position of $i(0)$, 
and the centre of the circle contain $R$ individuals, as observed in Fig. \ref{fig5}(b). For long times of simulation, 
the number of $R$ individuals increases until nearly the entire population 
(Fig. \ref{fig5}(c)). Due to permanent immunity, the number of infected 
individuals decreases exponentially.

\begin{table}[hbt]
    \centering
    \caption{Parameters of SIR for CA and ODEs. The initial 
    condition for $S$ is $S(0) = N - I(0)$.}
    \begin{tabular}{|c|c|c|}
    \hline
    Parameters & Values (CA) & Values (ODEs) \\
    \hline
    $N $ & 10000 & 1 \\
    $\beta$ & 0.2 & 0.2 \\
    $\gamma$ & 0.1 & 0.1\\
    $I(0)$  & 100 (0.01$N$)     & 0.01\\
    $R(0)$ & 0  & 0 \\
    \hline
    \end{tabular}
    \label{tab_4}
    \end{table}

We study the SIR model described by ODEs, that is given by three coupled differential equations:
\begin{eqnarray}
\frac{dS}{dt} & = & -\beta  \frac{S I} {N},  \label{sir_eq1}\\
\frac{dI}{dt} & = & \beta  \frac{S I} {N} - \gamma I, \label{sir_eq2}\\
\frac{dR}{dt} & = & \gamma I, \label{sir_eq3}
\end{eqnarray}
where $\beta$ is the infection rate and $\gamma$ is the recovery rate. 
This model also obeys the constraint $S + I + R = N$. This way, it is possible to
rewrite one of the Eqs. (\ref{sir_eq1})-(\ref{sir_eq3}) into an algebraic 
expression. For example, the system can be reduced to two differential 
equations by taking $R = N - S - I$. Due to the non-linearity in  
Eqs. (\ref{sir_eq1})-(\ref{sir_eq3}), it is not possible to reduce the number of equations as we made for the SI model. Another difference from SI model 
is that the transmission stops when there are no  
$I$ individuals, not when all $S$ becomes $I$. 

\begin{figure}[htb]
\centering
\includegraphics[scale = 0.2]{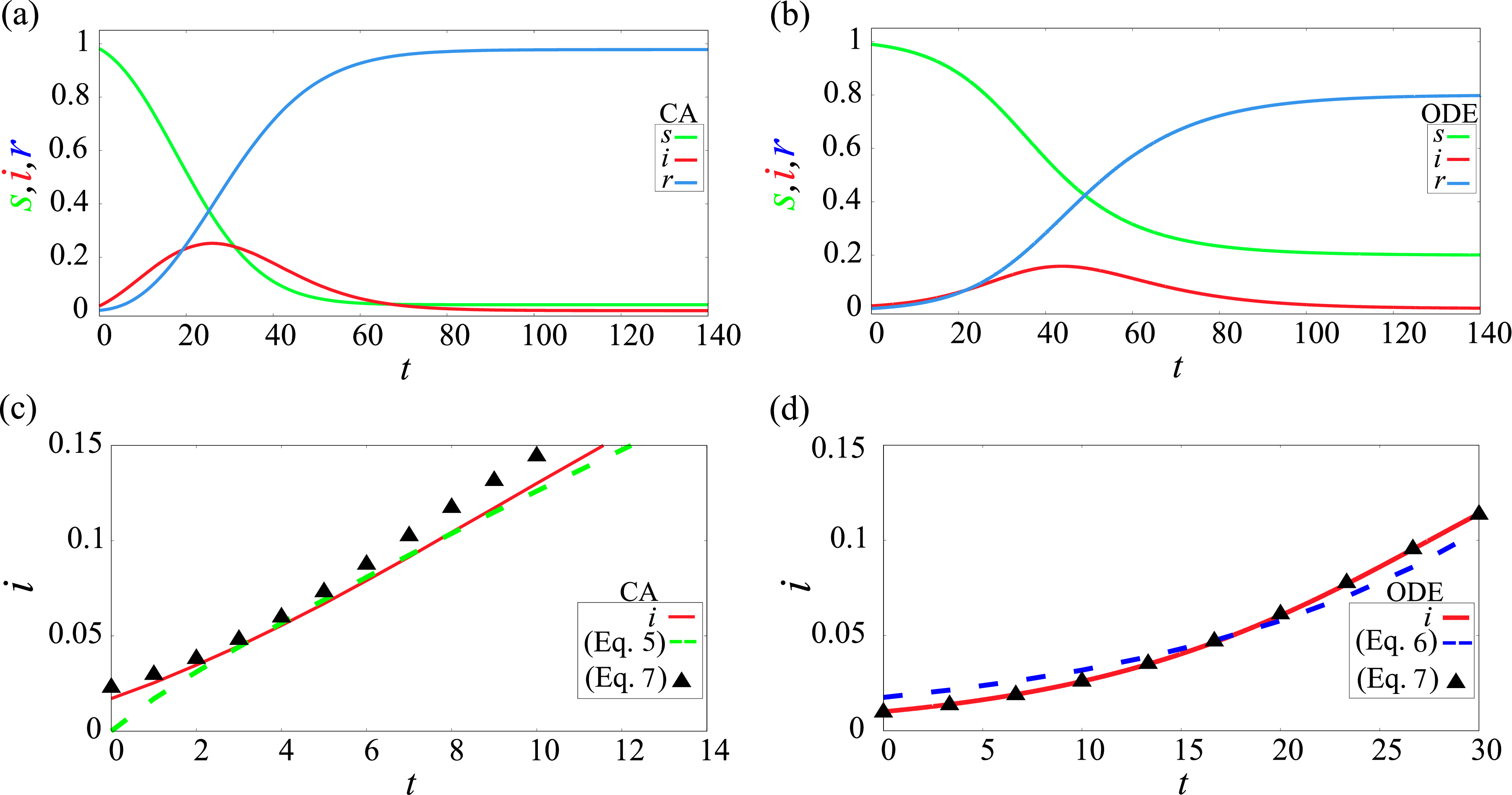}
\caption{Temporal evolution of $s$ (green line), $i$ (red line), and $r$ (blue line) 
of the SIR model for (a) CA and (b) ODE. 
In the panels (c) and (d), we show the $i$ curve for CA and ODE, 
respectively, followed by the adjustments 
given by Eq. (\ref{fit_CA}) in green dotted line,  Eq. (\ref{fit_ODE}) in blue dotted, and 
Eq. (\ref{fit_tanh}) by the black triangles. }
\label{fig6}
\end{figure}

A comparison between the $S$, $I$ and $R$ curves in the both frameworks is 
displayed in Fig. \ref{fig6}, where Fig. \ref{fig6}(a) is for CA and Fig.  
\ref{fig6}(b) for ODE. The behavior observed in these curves 
arises from the inherent nature of the models: CA captures  
spatial interactions, while ODE concentrates on temporal dynamics. 
CA allows the study of infection spread in spatially 
heterogeneous populations, revealing new patterns such as damped oscillations \cite{MISICI}. 
In contrast, the SIR model ODE focus on the temporal development of epidemics through non-linear 
differential equations. 

Another difference is in the growth of $I$ individuals, as noted previously. 
Following the same strategy, we fit the $I$ curve until the time where $I$ reaches 
$0.15N$. For 
CA, we consider Eq. (\ref{fit_CA}) and obtain parameters equal to 
$K = 0.017 \pm 0.001$, $B = 0.87 \pm 0.03$ 
and $R^2 = 0.9987$. This curve is displayed in Fig. \ref{fig6}(c) 
by the green dotted line. Meanwhile,  
for the ODE, we use Eq. (\ref{fit_ODE}) getting  
$K= 0.0175 \pm 0.0001$, 
$B = 0.0597 \pm 0.0001$, and $R^2 = 0.9926$ (blue dotted line in Fig. \ref{fig6}(d)). 
For the analysed case, we observe that the $I$ number 
in CA formalism grows like a power-law, while in ODE as an exponential 
function. We verify the adjustment given by Eq. (\ref{fit_tanh}), 
for CA: 
$\alpha = 0.090 \pm 0.003$, 
$\xi = 0.150 \pm 0.006$, 
$\phi = -1.030 \pm 0.019$, and 
$R^2 = 0.9996$; and
for the ODE: $\alpha = 0.10027 \pm 0.00007$, 
$\xi = 0.05447 \pm 0.00003$, $\phi = -1.5027 \pm 0.0003$, and 
$R^2 = 0.99996$. Both adjustments are shown in Figs. \ref{fig6}(c) 
and \ref{fig6}(d), respectively, by the black triangles. 

\section{SEIR model}\label{seir_section}
\begin{figure}[htb]
    \centering
    \includegraphics[scale = 0.14]{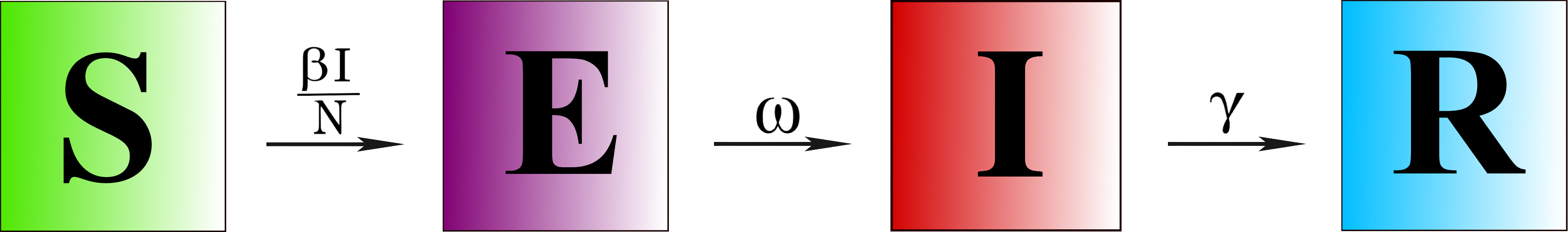}
        \caption{Compartment diagram for the SEIR model. 
	$\beta$ describes the effective rate of infection, 
	$\omega$ represents the rate of transfer from exposed to infectious, and 
	$\gamma$ represents the recovery rate.}
    \label{fig7}
\end{figure}

The SEIR model is represented by the diagram in Fig. \ref{fig7}, it is 
an extension of SIR by including a new compartment.  
This model is appropriate to describe the disease in which one
individual comes into contact with one infected agent and becomes contaminated. After that,  
the exposed individuals go through the latent period, during which they 
are not able to spread the disease. The time interval of latent period 
is equal to $1/\omega$. Once infected, the individual spreads the illness and after 
$1/\gamma$ becomes $R$. 

We analyze SEIR via CA, where  
$U = \{S,E,I,R\} = \{1,2,3,4\}$. In this model,  
$S$ cells become exposed if they are adjacent to infected cells and  
the transition is determined by the probability $\beta$.  Exposed cells progress 
to the $I$ state with probability $\omega$, that is related to the  
incubation period. During the incubation period the exposed cell does not 
transmit the disease to its neighbours. Infected cells recover with a 
probability $\gamma$, transitioning to the $R$ state. 
The CA rule for the SEIR model 
can be summarised as follows
\begin{enumerate}
	\item The evolution of the $S$ cell to E state 
	occurs with a probability $\beta$ (infection rate), if there is at 
	least one adjacent infected cell.
	\item Exposed cells do not transmit the diseases 
	to other cells. These cells can transition to infected state 
	($C_{i,j}(t) = 3$) with probability $\omega$ (incubation rate).
	Once the exposed cell evolves to an infected cell, it begins to 
	expose adjacent cells to the pathology.
	\item An infected cell can recover from the diseases 
	with probability $\gamma$ (recovery rate). Once recovered, the infected 
	cell updates to a recovered cell, 
	state which the cell remains for the entire simulation. 
\end{enumerate}

\begin{figure}[htb!]
\centering
\includegraphics[scale = 0.38]{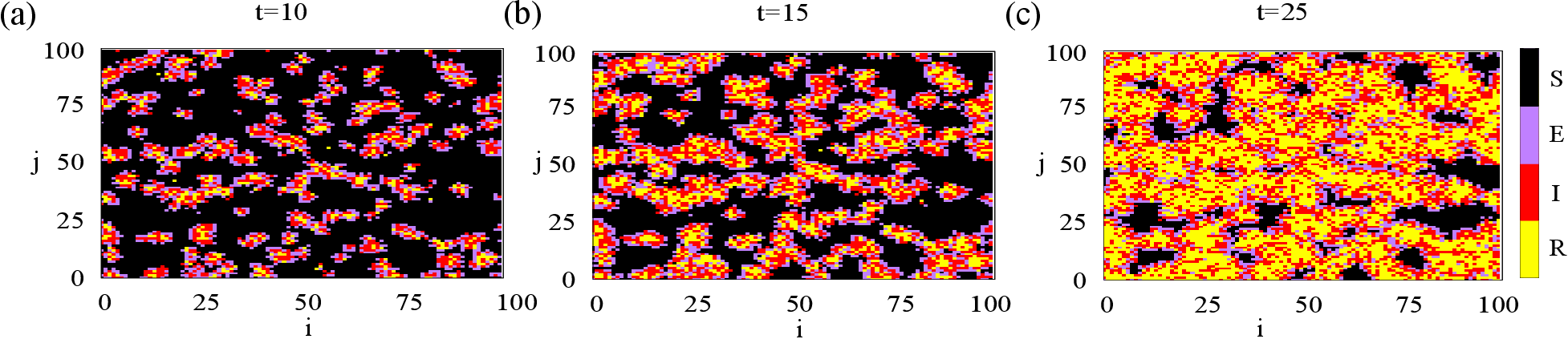}
\caption{Spatial-temporal evolution of the CA for SEIR for (a) $t=10$, (b) $t=15$, and (c) $t=25$. 
The colour of each individual represents the compartment of the cell:
         susceptible individuals are represented by black, exposed by purple, infected by red and recovered by yellow.}
\label{fig8}
\end{figure}

Implementing the previous rules, we obtain the spatial-temporal evolution 
for the SEIR model, as exhibited in Fig. \ref{fig8}(a)-(c), for $t=10$, 15, and 25, 
respectively. The parameters are displayed  
in Table \ref{tab_6}. As $I$ advances infecting 
individuals, the agents in the centre of the circle evolve to $R$. The final 
state is a grid full of $R$.
\begin{table}[hbt]
    \centering
    \caption{Parameters considered in the CA for SEIR. The initial 
    condition for $S$ is $S(0) = N - I(0) - E(0)$.}
    \begin{tabular}{|c|c|c|}
    \hline
    Parameters & Values (CA) & Values (ODE) \\
    \hline
    $N$ & 10000 & 1\\
    $\beta$ & 0.25 & 0.25\\
    $\gamma$ & 0.1 & 0.1\\
    $\omega$ & 0.2 & 0.2\\
	$E(0)$ & 100 (0.01$N$) & 0.01\\
    $I(0)$ & 100 (0.01$N$)  & 0.01\\ 
    $R(0)$ & 0  & 0\\
    \hline
    \end{tabular}
    \label{tab_6}
    \end{table}

Considering an ODE approach, the SEIR model is described by
\begin{eqnarray}
\frac{dS}{dt} & = & \frac {-\beta S I} {N}, \label{seir_eq1} \\
\frac{dE}{dt} & = & \frac {\beta S I} {N} - \omega E, \label{seir_eq2} \\
\frac{dI}{dt} & = & \omega E - \gamma I, \label{seir_eq3} \\
\frac{dR}{dt} & = & \gamma I, \label{seir_eq4}
\end{eqnarray}
where the restriction $N = S + E + I +R$ is valid. Using the constraint, 
the Eqs. (\ref{seir_eq1})-(\ref{seir_eq4}) are reduced to three ODEs 
systems by computing $R$ via $R = N - (S + E + I)$.
The dynamics are more complex than those of the SIR 
model due to the inclusion of the exposed compartment, 
which represents a delay between infection and the onset of infectiousness. 
The delay can influence both the speed and intensity of epidemic outbreaks. 
The disease can spread more slowly initially due to the incubation period, 
however, once the exposed individuals become infectious, there can be a rapid 
increase in the cases. 

\begin{figure}[htb]
\centering
\includegraphics[scale = 0.2]{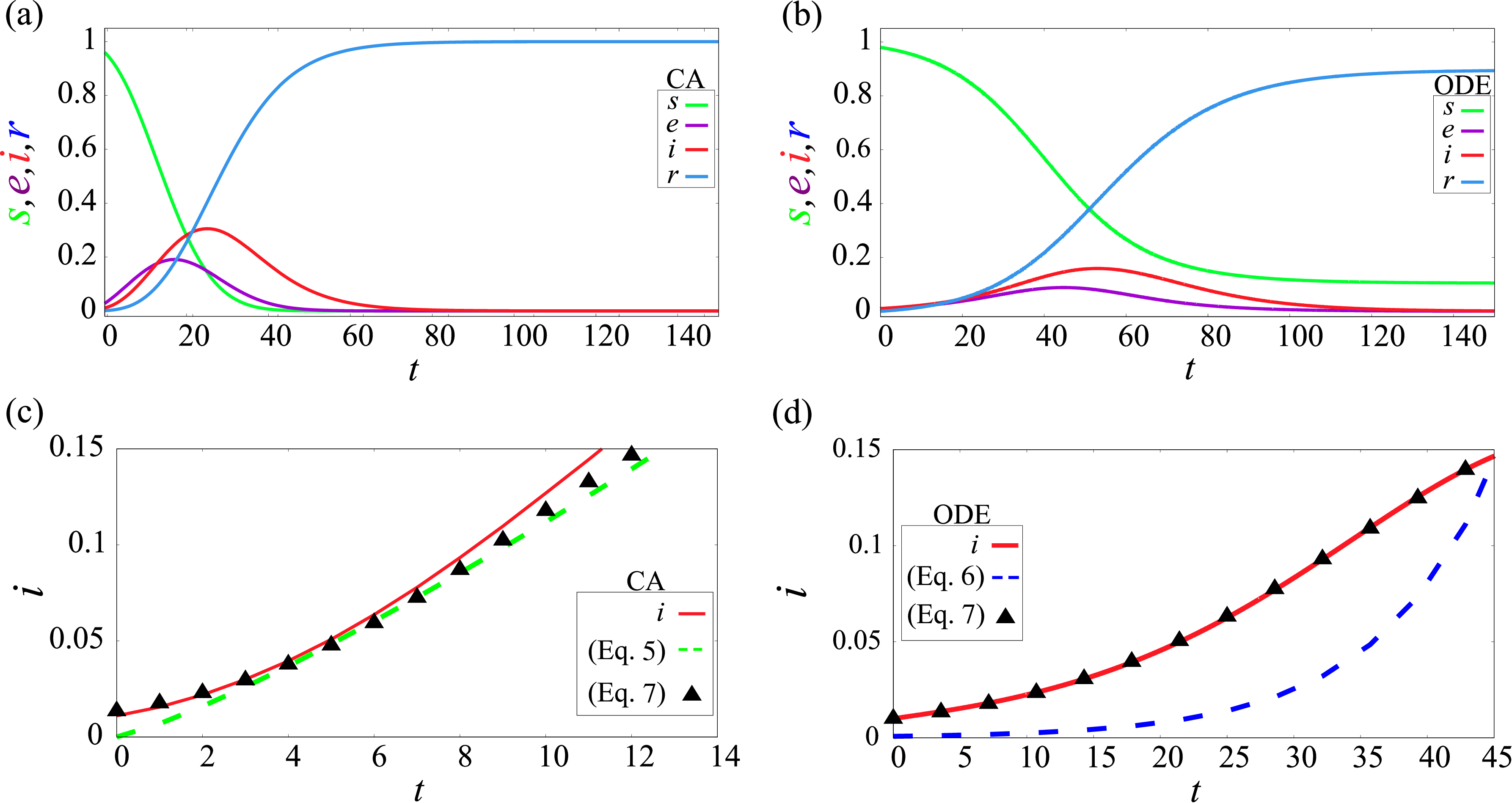}
\caption{Temporal evolution of $s$ (green line), $e$ (purple line), $i$ (red line), and $r$ (blue line) 
of SEIR for (a) CA and (b) ODE. In panels (c) and (d), we show the $i$ curve for CA and ODE, 
respectively, followed by the adjustments 
given by Eq. (\ref{fit_CA}) in green dotted line,  Eq. (\ref{fit_ODE}) blue dotted, and 
Eq. (\ref{fit_tanh}) by the black triangles. 
         }
\label{fig9}
\end{figure}

The temporal evolution of $S$, $E$, $I$, and $R$ individuals 
are shown in Fig. \ref{fig9}. The panels (a) and (b) shows the temporal 
evolution for CA and ODEs formalism, respectively. The curve shape of 
CA and ODE approach are similar. 
The difference occurs basically in the beginning of the outbreak. Note that the 
E individuals increase faster in CA, while in ODE, they grow slowly 
for $t<20$ and after that assume an exponential increase. 

To verify which function better describes the start of the outbreak. We  
adjust the CA curve via Eq. (\ref{fit_CA}), getting 
$K = 0.0069 \pm 0.0005$, 
$B = 1.21 \pm 0.03$, and 
$R^2 = 0.9987$ (green dotted line in Fig. \ref{fig9}(c)). 
For the ODE approach, via Eq. (\ref{fit_ODE}), we obtain  
$K = 0.00077 \pm 0.00008$, $B = 0.116 \pm 0.002$, and $R^2 = 0.7509$ 
(blue dotted line in Fig. \ref{fig9}(d)).  
Then, we employ Eq. (\ref{fit_tanh}) for CA and ODE. For 
the first one, the parameter is 
$\alpha = 0.106 \pm 0.003$, 
$\xi = 0.142 \pm 0.004 $,
$\phi = -1.35 \pm 0.01$, and 
$R^2 = 0.9997$; and for the second
$\alpha = 0.10360 \pm 0.00007$, 
$\xi = 0.04334 \pm 0.00002$, 
$\phi = -1.4970 \pm 0.0003$, and
$R^2 = 0.99996$. The black triangles display both fittings in 
Fig. \ref{fig9}(c) and (d). Comparing the previous descriptions for SI 
and SIR, the power-law and exponential do not describe very well the 
$i$ curve for SEIR, otherwise, the hyperbolic tangent keeps being a 
good adjustment. 

\section{Conclusions}\label{discussion_section}
In this work, we investigate three epidemiological models: 
SI, SIR and SEIR. To simulate these models, we use two different techniques:
cellular automata (CA) and ordinary differential equations (ODE). We made 
a direct comparison between both approaches. Our results shows that the 
mathematical representation affects profoundly the modelling output. 
Our main observation is that the infected initial growth in CA approach is better 
described by a power-law, while in ODE the infected number grows as an 
exponential function. In addition, we propose an adjustment 
based on hyperbolic tangent, which represents very well both growth.  
It is important 
to know how $I$ grows in the beginning of outbreak. 
Knowing this, we are able to select the mathematical description 
discussed in this paper. 

{ 
An important contribution of our work is the use of a hyperbolic tangent to fit the initial growth of all models in both approaches. The use of hyperbolic tangent is presented in \cite{KEELING} where the authors obtain an approximate solution for $R(t)$ in the limit when $R_0 R$ is small for the SIR model. This condition, sometimes, coincides with the initial growth of outbreaks. Based on this, we show via numerical tests that a hyperbolic tangent can fit not only the SIR model, but also SI and SEIR. We address these successful results by analyzing the representation of the hyperbolic tangent in terms of infinite series. We observe a sum of alternating signs of odd powers describes it. This suggests that the hyperbolic tangent better captures the growth when compared to a power-law representation that uses a single exponent. Additionally, the hyperbolic tangent is defined by exponential functions that preserves a relationship with the EDOs solutions. This combination of characteristics makes the hyperbolic tangent more suitable for simultaneously representing both CA and EDO. }

The comparison between CA and ODE reveals that both methods offer 
distinct perspectives for modelling and analysing epidemics, each with 
its own advantages and limitations. CA, with their discrete approach and 
ability to simulate local interactions between individuals, provides a 
detailed view of disease spread within a spatially organized structure.  
Meanwhile, ODE with their continuous and analytical approach, offers a 
simplified and effective overview of the trends and rates of 
disease spread in large and homogeneous populations. The ability of ODEs 
to generate analytical or numerical solutions for dynamic systems permits  
quantitative analysis and the prediction of long-term epidemiological 
behaviors. To summarise the comparison, we say that CA model described 
a microscopic level, i.e., individual by individual, while ODE gives the 
macroscopic description. 

The use of CA and ODE in epidemiology expands the 
scope of possible analyses and offers complementary tools that help build 
a more detailed picture of disease propagation dynamics. Our 
study shed light into the differences between ODE and CA compartmental 
epidemiological models. Different mathematical approaches can better describe 
the experimental data of specific diseases. Therefore, combining these 
methodologies can enhance intervention and control strategies, promoting 
more effective outbreak and pandemic management. { In this way}, continuous research 
and development of these techniques are essential to tackle emerging 
epidemiological challenges and improve public health policies based on solid evidence.


\section*{Acknowledgements}
We acknowledge CAPES (Financial Codes: 88887.991936/2024--00, 88887.650484/2021--00, 
88887.856951/2023--00, 88887.965333/2024-00), CNPq (Fincancial Code:  309670/2023-3), Funda\c{c}\~ao Arauc\'aria, 
and FAPESP (Financial code: 2024/14478-4, 2024/14825-6,  	
2025/02318-5). 
We also would 
like to acknowledge the Science Group 105 (www.105groupscience.com)

\section*{Funding}
The authors would like to thank the Brazilian funding agencies: CAPES (Financial Codes: 88887.991936/2024--00, 88887.650484/2021--00, 
88887.856951/2023--00, 88887.965333/2024-00), CNPq (Fincancial Code:  309670/2023-3), Funda\c{c}\~ao Arauc\'aria, 
and FAPESP (Financial code: 2024/14478-4, 2024/14825-6,  	
2025/02318-5) by the financial support.

\section*{Data Availability}
The numerical data that support the findings of this study are
available from the corresponding author upon reasonable
request.

\section*{Declarations}
{\textbf Conflict of Interest:}\\
The author declares that there exists no competing financial interest or 
personal relationships that could have appeared to influence the work 
reported in this paper.


\end{document}